\documentclass[usenatbib]{mnras}

\usepackage{graphicx}
\usepackage{natbib}
\usepackage{rotating}
\usepackage{epsfig}
\usepackage{aas_macros}
\usepackage[toc,page]{appendix}
\usepackage{multirow}
\usepackage{setspace}
\usepackage{placeins}
\usepackage{multicol}
\usepackage{float}
\usepackage{epsf}
\usepackage{tabularx}
\usepackage{multirow}
\usepackage{balance}
\usepackage{amsmath}
\usepackage{latexsym}
\usepackage{url}
\usepackage{bm}
\usepackage{stmaryrd,xspace}

\usepackage{amsmath}

\newcommand{\Ttot}{\ensuremath{T_{\text{tot}}}}
\newcommand{\Tce}{\ensuremath{T_{\text{ce}}}}
\newcommand{\Ltot}{\ensuremath{L_{\text{tot}}}}
\newcommand{\Lce}{\ensuremath{L_{\text{ce}}}}
\newcommand{\Lcxo}{\ensuremath{L_{\text{52}}}}
\newcommand{\Lcxoi}{\ensuremath{\hat{L}_{\text{52,i}}}}
\newcommand{\Lcxoh}{\ensuremath{\hat{L}_{\text{52}}}}
\newcommand{\Lrosat}{\ensuremath{L_\text{400d}}}
\newcommand{\Lrosati}{\ensuremath{\hat{L}_{\text{400d,i}}}}
\newcommand{\Lrosath}{\ensuremath{\hat{L}_{\text{400d}}}}

\newcommand{\rf}{\ensuremath{R_{\text{500}}}}

\newcommand{\lik}{\ensuremath{{\cal L}}}

\newcommand{\etal}{et al.\ }

\newcommand{\Chandra}{\emph{Chandra}}

\newcommand{\ROSAT}{\emph{ROSAT}}
\newcommand{\XMM}{\emph{XMM-Newton}}

\newcommand{\chisq}{\ensuremath{\chi^2}}

\newcommand{\gta}{\,\rlap{\raise 0.4ex\hbox{$>$}}{\lower 0.6ex\hbox{$\sim$}}\,}
\newcommand{\lta}{\,\rlap{\raise 0.4ex\hbox{$<$}}{\lower 0.6ex\hbox{$\sim$}}\,}

\newcommand{\cm}{\mbox{\ensuremath{\text{~cm}}}}

\newcommand{\km}{\mbox{\ensuremath{\text{~km}}}}

\newcommand{\Mpc}{\mbox{\ensuremath{\text{~Mpc}}}}
\newcommand{\s}{\mbox{\ensuremath{\text{~s}}}}

\newcommand{\keV}{\mbox{\ensuremath{\text{~keV}}}}

\newcommand{\erg}{\mbox{\ensuremath{\text{~erg}}}}

\newcommand{\pcmsq}{\ensuremath{\cm^{-2}}}
\newcommand{\pMpc}{\ensuremath{{\Mpc^{-1}}}}
\newcommand{\ps}{\ensuremath{\s^{-1}}}

\newcommand{\ergps}{\ensuremath{{\erg \ps}}}
\newcommand{\flux}{\ensuremath{\erg \ps \pcmsq}}

\newcommand{\kmpspMpc}{\ensuremath{{\km \ps \pMpc\,}}}

\newcommand{\apropto}{\mathrel{\vcenter{
  \offinterlineskip\halign{\hfil$##$\cr
    \propto\cr\noalign{\kern2pt}\sim\cr\noalign{\kern-2pt}}}}}

\newcommand{\LT}{\mbox{\ensuremath{L-T}}}

\newcommand{\YM}{\mbox{\ensuremath{Y_{X}-M_{500}}}}
\newcommand{\LM}{\mbox{\ensuremath{L-M}}}

\newcommand{\MT}{\mbox{\ensuremath{M-T}}}

\newcommand{\ALT}{\ensuremath{A_{LT}}}
\newcommand{\BLT}{\ensuremath{B_{LT}}}
\newcommand{\gLT}{\ensuremath{\gamma_{LT}}}
\newcommand{\intLT}{\ensuremath{\delta_{LT}}}
\newcommand{\xcal}{\ensuremath{X_\text{cal}}}

\def\onecolumn{%
  \global\columnwidth\textwidth
  \global\hsize\columnwidth
  \global\linewidth\columnwidth
  }

\title[The L-T relation of low mass galaxy clusters]{The X-ray luminosity temperature relation of a complete sample of low mass galaxy clusters}
\author[S. Zou \etal]{
\parbox[h]{\textwidth}{S. Zou,$^{1,2}$ B. J. Maughan,$^1$ P. A. Giles,$^1$
  A. Vikhlinin,$^{3,4}$ F. Pacaud,$^5$  R. Burenin,$^{4,6}$ and A. Hornstrup$^7$
}
\vspace*{12pt} \\
\parbox[h]{\textwidth}{
$^1$HH Wills Physics Laboratory, University of Bristol, Tyndall
  Avenue, Bristol, BS8 1TL, UK\\
$^2$Institut d'Astrophysique de Paris, Universit\'e Paris 6, CNRS-UMR7095,
 98bis Boulevard Arago, 75014 Paris, France\\
$^3$Harvard-Smithsonian Center for Astrophysics, 60 Garden Street, Cambridge, MA 02140, USA\\
$^4$Space Research Institute, Russian Academy of Sciences, Profsoyuznaya ul. 84/32, Moscow, 117997, Russia\\
$^5$Argelander-Institut fur Astronomie, Univeristy of Bonn, Auf dem Hugel 71, D-53121 Bonn, Germany\\
$^6$Moscow Institute of Physics and Technology, Dolgoprudny, Institutsky per., 9, 141700, Russia\\
$^7$National Space Institute, Technical University of Denmark, Juliane Maries Vej 30, 2100 Copenhagen 0, Denmark\\
 }}
\date{Accepted 2016 August 8. Received 2016 August 3; in original form 2015 November 23}
\begin{document}

\maketitle
\begin{abstract}

  We present \Chandra\ observations of 23 galaxy groups and low-mass
  galaxy clusters at $0.03<z<0.15$ with a median temperature of
  $\sim2\keV$. The sample is a statistically complete flux-limited
  subset of the 400 deg$^2$ survey. We investigated the scaling
  relation between X-ray luminosity ($L$) and temperature ($T$),
  taking selection biases fully into account. The logarithmic slope of
  the bolometric \LT\ relation was found to be $3.29\pm0.33$,
  consistent with values typically found for samples of more massive
  clusters. In combination with other recent studies of the \LT\
  relation we show that there is no evidence for the slope,
  normalisation, or scatter of the \LT\ relation of galaxy groups
  being different than that of massive clusters. The exception to this
  is that in the special case of the most relaxed systems, the slope
  of the core-excised \LT\ relation appears to steepen from the
  self-similar value found for massive clusters to a steeper slope for
  the lower mass sample studied here. Thanks to our rigorous treatment
  of selection biases, these measurements provide a robust reference
  against which to compare predictions of models of the impact of
  feedback on the X-ray properties of galaxy groups.

\end{abstract}


\section{Introduction}
Galaxy clusters are the largest gravitationally bound systems in the
Universe, ranging in size from 2 -- 10 Mpc, with X-ray luminosities of
$\sim 10^{43} - 10^{45} $ erg s$^{-1}$. The mass content of clusters
consists of $\sim85\%$ dark matter, $\sim12 \%$ X-ray bright,
low-density intra-cluster medium (ICM), and $\sim3 \%$ stars. Studying
galaxy clusters is motivated by two complementary goals, investigating
the formation and evolution of clusters and their galaxies, and using
clusters as cosmological probes.

If the ICM is only heated by the conversion, via shocks, of its
gravitational potential energy to internal energy during its infall
into the cluster, then its properties will exhibit self-similar
behaviour. This will lead to simple power-law correlations between the
X-ray observables, such as the temperature ($T$) and luminosity ($L$)
of the gas \citep{kai6}. Importantly, any deviations of observed
clusters from this self-similar behaviour points to the action of
non-gravitational energy input to the ICM, such as mechanical and
radiative energy from supernova-driven galaxy winds, or outflows
powered by active galactic nuclei (AGN).

The correlation between X-ray luminosity and temperature (the \LT\
relation) has been extensively studied, due to the relative ease with
which those properties can be measured
\citep[e.g.][]{edg91,mar98,pra09,ben12,lov14,bha14a}. A consensus has
emerged that the \LT\ relation is steeper than the self-similar
prediction, in the sense that lower mass clusters are hotter and/or
less luminous than expected (although \cite{ben12} found
  evidence that the most massive, relaxed clusters show self-similar
  behaviour when their core regions are ignored). This is interpreted
as evidence that non-gravitational heating has a stronger impact on
the ICM in low mass halos where the gravitational potential is weaker,
leading to similarity breaking.

Further evidence for the presence of non-gravitational heating in
groups and clusters is provided by observations of cluster cores. In
core regions, the high ICM density leads to cooling times that are
short relative to the cluster's lifetime. This should establish a
cooling flow, wherein cooling, condensing gas in the cluster core is
replaced by a slow inflow of gas from larger radii, which itself cools
as it flows into the core \citep[see ][ for a review]{fab94}. The high
cooling rates expected in this scenario have not been observed, with
observations demonstrating that the ICM in these cool cores is being
prevented from cooling fully out of the X-ray emitting regime in large
quantities \citep[e.g.][]{pet06}.

The favoured mechanism for balancing cooling in cluster cores is
energy input from AGN, based on evidence including the large fraction
of cool-core clusters that possess AGN which show signs of interacting
with the ICM by blowing cavities, and plausibility arguments that the
energy associated with these cavities (and in some cases, related
shocks) is sufficient to balance cooling
\citep[see e.g.][]{chu02,mcn07,pan14b,ran15}.

It is thus plausible that AGN input is responsible for both breaking
self-similarity in clusters, and balancing cooling in their cores. An
emerging model is that two modes of AGN feedback are at work
\citep[as reviewed in][]{feedbackreview}. Cooling appears to be balanced
by ongoing mechanical energy input from the AGN, forming a feedback
loop with the accretion of cooling gas onto the central galaxy. It is
proposed that self-similarity was broken by a form of AGN heating that
raised the entropy of the gas in clusters, reducing its density (and
hence X-ray luminosity) and removing it towards or beyond the virial
radius of the cluster. It is unclear if this heating occurred in the
form of energy input through winds or outbursts at high redshifts, or
steady continuous feedback over time. It is also possible that excess
energy from the ongoing mechanical feedback, beyond that required to
balance cooling, could play a role in similarity breaking over the
lifetime of the cluster \citep{hla14}. Studying the ICM properties of
cluster populations as a function of mass and redshift can
discriminate between, and refine these feedback models.

The study of the \LT\ relation in low-mass clusters and galaxy groups
thus has the potential to give clues to the nature of the
non-gravitational processes that break self-similarity. However,
measuring the \LT\ relation in groups is more challenging than for
clusters because of their lower intrinsic luminosities. Progress has
been made, albeit without yet reaching the same level of consensus
that is found in the higher mass regime. Earlier studies have
variously found that the \LT\ relation in groups is consistent with
\citep[e.g.][]{mul98,osm04} or steeper than \citep[e.g.][]{hel00} that
in clusters.

Some of the variety in results may be due to the composition of the
samples that have been studied. Groups and clusters undergoing mergers
trace stochastic paths on the \LT\ plane as the merger progresses
\citep{row04}. Meanwhile, the presence of cool cores in systems leads
to large offsets in the \LT\ plane relative to those without cool
cores \citep[e.g.][]{mar98}. These effects contribute to significant
scatter in the \LT\ plane, and can lead to disparate results if
non-representative samples of clusters are studied. More recently, the
importance of selection biases on the determinations of cluster
scaling relations has been recognised, which could have significant
effects in the low-mass regime, where the scatter in observables is
expected to be large. Studies that have attempted to correct for
selection biases on the \LT\ relation appear to be converging to show
that the slope of the \LT\ relation is consistent between clusters and
groups \citep{man10,lov14,bha14a}.

In this paper we investigate the \LT\ relation of a complete sample of
groups and low-mass clusters selected from the 400 square degree
survey \citep[400d; ][]{bur07}. Our work is comparable with other
recent studies of the \LT\ relation in this low mass regime
\citep{lov14,bha14a}, but we employ what is arguably the most rigorous
treatment of selection biases used in this mass range, comparable with
the approach of \citet{man10} for higher-mass clusters. The
  sample studied in \citet[][B14 hereafter]{bha14a} comprised 26
  groups taken from several cluster surveys based on ROSAT all-sky
  survey (RASS) data, and that had available \Chandra\ observations
  \cite[as described in][]{eck11}. The selected groups have a
  redshift range $0.01<z<0.05$, and temperatures in the range
  $0.6\leq T \leq3.6 \keV$. The work of \citet[][L14 hereafter]{lov14}
  is based on a sample selected in a similar way from RASS-based
  surveys, but they were able to construct a complete flux-limited
  sample of 23 clusters with \XMM\ observations. The resulting sample
  had a redshift range $0.01<z< 0.04$ and temperature range
  $0.85\leq T \leq 2.80 \keV$. The B14 and L14 samples have 8 groups
  in common with each other, but neither overlap with our 400d groups
  sample.

This paper is organized as follows: we describe the sample and its
analysis in \textsection \ref{sec:analysis}; we present \LT\ relation
both with and without correction for selection biases in \textsection
\ref{sec:bces} and \textsection \ref{sec:lt-relation-with}; in
\textsection \ref{sec:discussion} we discuss our results and compare
them with other recent work; finally, the key results are summarised
in \textsection \ref{sec:summary}. We use a standard $\Lambda$CDM
cosmology throughout this paper, with $\Omega_{\Lambda}$ = 0.7,
$\Omega_M$ = 0.3, and $H_0=70\kmpspMpc\equiv 100 h \kmpspMpc$. The
function $E(z)$ arises in the evolution of the scaling relations and
describes the evolution of the Hubble parameter. It is given by
$E(z) = \sqrt{\Omega_{M}(1+z)^3 + \Omega_\Lambda}$.

\section{Data Analysis}
\label{sec:analysis}
\begin{table*}
  \caption{\label{sample} Summary of the 400d groups sample and the
    \Chandra\ data used. RA and DEC and redshift information are from
    \citet{bur07}. The \Chandra\ observation ID and detector (ACIS-I
    or ACIS-S) are given, along with the cleaned exposure time of the
    \Chandra\ observation. Column 8 gives a number indicating the
      paper presenting the original analysis of the observations (where available) as
      follows: (1)\citet{sun09a}; (2)\citet{sun09}; (3) \citet{ma11}; (4)\citet{hof16} ; (5) \citet{vik05};
    (6)\citet{kno07} }
\centering
\begin{tabular}{lccccccll}
\hline
 Clusters         &RA               &   DEC        & Redshift
  & Observation ID      &ACIS-I/S            &Exposure (ks)& PI & Reference\\
\hline
 Cl0327+0233    &03:27:54.5  & +02:33:47 &  0.030 &    9391  &   I &      11.12 & Vikhlinin   & 1\\
 Cl0306-0943    &03:06:28.7  &-09:43:50  & 0.034  &    9389  &   I &      10.05 & Vikhlinin   &  1\\
 Cl1058+0136    &10:58:12.6  &+01:36:58  & 0.038  &    9387  &   I &      10.05 & Vikhlinin   &  2\\
 Cl1259+3120    &12:59:51:0  &+31:20:48  & 0.052  &    9395  &   I &      17.92 & Vikhlinin   &  1\\
 Cl0334-3900    &03:34:03.3  &-39:00:46  &0.062   &    9393  &   I &      15.56 & Vikhlinin   &  1\\
 Cl0810+4216    &08:10:24.2  &+42:16:19  &0.064   &    13986 &   I &       9.95 & Maughan     & \\
 Cl1630+2434    &16:30:14.7  &+24:34:47  & 0.065  &    9386  &   I &       9.65 & Vikhlinin   &  \\
 Cl1533+3108    &15:33:17.1  &+31:08:55  &  0.067 &    9384  &   I &       9.99 & Vikhlinin   &  2\\
 Cl0340-2840    &03:40:27.2  &-28:40:20  & 0.068  &    9385  &   I &       9.64 & Vikhlinin   &  \\
 Cl1206-0744    &12:06:33.5  &-07:44:24  &  0.068 &    9388  &   I &      10.01 & Vikhlinin   &  2\\
 A1775          &13:41:52.0  &+26:22:49  & 0.076  &    12891 &   S &      39.52 & Gastaldello &4\\
   -            & -          & -         &  -     &    13510 &   S &      59.26 & Gastaldello &4\\
 A744           &09:07:20.0  &+16:39:25  &  0.076 &    6947  &   I &      39.52 & Vikhlinin   &  2\\
 RXJ1159+5531   &11:59:51.2  &+55:31:56  & 0.081  &    4964  &   S &      75.11 & Buote   &      5\\
 Cl2220-5228    &22:20:09.1  &-52:28:01  &  0.102 &    9383  &   I &      10.04 & Vikhlinin   &   \\
 Cl0336-2804    &03:36:49.4  &-28:04:53  & 0.104  &    9390  &   I &      10.63 & Vikhlinin   &   3\\
 Cl1501-0830    &15:01:18.3  &-08:30:33  &  0.108 &    13987 &   I &       9.94 & Maughan     &    \\
 A2220          &16:39:55.5  &+53:47:55  & 0.111  &    7876  &   S &      14.95 & Jetha       &      1\\
    -           & -          &  -        & -      &    9832  &   S &      18.81 & Jetha       &      1\\
 Cl0057-2616    &00:57:24.7  & -26:16:49 & 0.113  &    9427  &   I &       9.99 & Vikhlinin   &  3\\
 Cl0838+1948    &08:38:31.4  &+19:48:15  & 0.123  &    9397  &   I &      19.97 & Vikhlinin   &   3\\
 Cl0237-5224    &02:37:59.6  &-52:24:47  & 0.134  &    9392  &   I &      13.95 & Vikhlinin   &  \\
 Cl1552+2013    & 15:52:12.3 &+20:13:42  & 0.136  &    3214  &   S &      14.93 & Jones       &    6\\
 RXJ1416.4+2315 &14:16:26.8  &+23:15:30  & 0.138  &    2024  &   S &      14.57 & Jones       &    6\\
 Cl0245+0936    &02:45:45.7  & +09:36:36 &0.147   &    9394  &   I &      14.97 & Vikhlinin   &  3\\

\hline
\end{tabular}
\end{table*}

\begin{figure}
\begin{center}
\hspace{-0.75cm}\includegraphics[width=9.1cm]{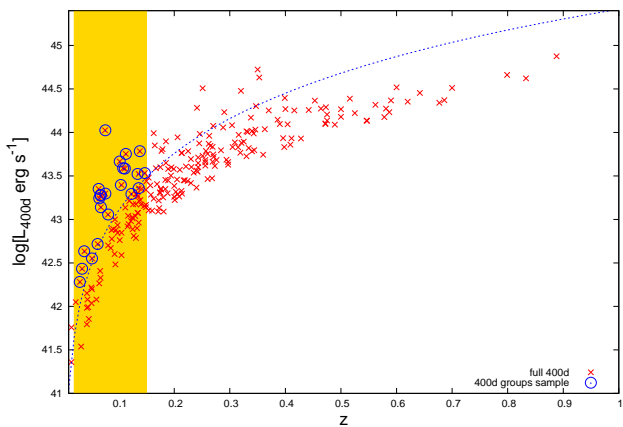}
\end{center}
\caption{\label{limit}Distribution of the 400d clusters in the
  luminosity-redshift plane. Red crosses show the full 400d
  survey, blue circles within the golden box show the 400d groups
    sample studied in this work, with the gold box highlighting the
  0.03$<$z$<$0.15 range used to select the clusters. The blue dashed
  line give the flux limit of the 400d groups sample of
  $5\times 10^{-13}\flux$.}
\end{figure}

\subsection{Sample selection}
The sample used in this work is a subset of 23 low-mass clusters or
groups from the 400d survey. The 400d survey \citep{bur07} is an
extension of 160d ROSAT PSPC survey \citep{vik98}, and is based on 1610
pointings covering an area of 397 deg$^{2}$, using the same detection
algorithm as the 160d survey. Observations of the evolution of the
clusters were used to place tight constraints on cosmological
parameters \citep{vik09,vik09a}. Our subsample is complete
above a flux limit of 5 $\times 10^{-13}\flux$, and within a redshift
range $0.03<z<0.15$. The flux limit is defined in the observer's frame
0.5-2 keV energy band from the \ROSAT\ observations. The clusters have
all been re-observed with \Chandra, and the details of the \Chandra\
observations are provided in Table \ref{sample}. Figure \ref{limit}
shows the luminosity-redshift distribution in the 400d survey, and
highlights the subsample used for the current work. The sample
comprises galaxy groups and low-mass clusters, with a median
temperature of $\sim2\keV$, and for convenience it is referred to as
the 400d groups sample.

\subsection{Data reduction and analysis}
The \Chandra\ observations of the clusters were reduced and analysed
using CIAO \citep{fru06}version 4.6.1.
The data were reprocessed from level 1 events using the \Chandra\
calibration database \citep{fru06}
version 4.6.1, and the normal data cleaning and reduction steps were
followed. In particular, background light curves were produced and
cleaned to remove periods of high background in a manner consistent
with the blank-sky background data sets, which were subsequently used
to estimate the background for our spectral analyses. Most of the
clusters in this work were observed in VFAINT mode, so the additional
VFAINT cleaning process was applied where appropriate.

\setlength{\extrarowheight}{.4em}
\begin{table*}
  \caption{\label{ltc} Cluster properties measured in the $(0-1)\rf$
    aperture. \Ltot\ and \Lcxo\ are the \Chandra\ luminosities in the
    bolometric and soft ($0.5-2\keV$) band. \Lrosat\ is the ROSAT PSPC
    luminosity in the same soft band from \citep{bur07}; unlike the properties
    measured with \Chandra\, this is the total luminosity of the cluster, not the
    luminosity within $\rf$.}
\centering
\begin{tabular}{lcccccc}
\hline
Cluster &       $z$ & $\rf$  & $\Ttot$  & $\Ltot$  & $\Lcxo$ & $\Lrosat$ \\
  &        & (Mpc) & (keV) & ($10^{43}\ergps$) & ($10^{43}\ergps$) & ($10^{43}\ergps$) \\
\hline
Cl0327+0233   &       0.030& 0.374&	$0.98^{+0.04}_{-0.04}$     & 0.36$\pm$0.02 & 0.20$\pm$0.01 & 0.19$\pm$0.02 \\
Cl0306-0943   &       0.034& 0.411&	$1.15^{+0.06}_{-0.07}$     & 0.68$\pm$0.08 & 0.37$\pm$0.05 & 0.27$\pm$0.03 \\
Cl1058+0136   &       0.038& 0.559&	$2.30^{+0.29}_{-0.25}$     &10.70$\pm$0.90 & 0.56$\pm$0.05 & 0.43$\pm$0.06 \\
Cl1259+3120   &       0.052& 0.416&	$1.05^{+0.11}_{-0.08}$     & 0.55$\pm$0.85 & 0.29$\pm$0.04 & 0.36$\pm$0.10 \\
Cl0334-3900   &       0.062& 0.666&	$2.53^{+0.59}_{-0.46}$     & 2.01$\pm$0.18 & 0.83$\pm$0.07 & 0.58$\pm$0.04 \\
Cl0810+4216   &       0.064& 0.777&	$3.51^{+0.41}_{-0.50}$     & 6.63$\pm$0.43 & 2.86$\pm$0.16 & 2.24$\pm$0.25 \\
Cl1630+2434   &       0.065& 0.730&	$2.86^{+0.44}_{-0.43}$     & 5.10$\pm$0.43 & 2.09$\pm$0.18 & 1.75$\pm$0.25 \\
Cl1533+3108   &       0.067& 0.608&	$1.67^{+0.12}_{-0.11}$     & 3.35$\pm$0.26 & 1.66$\pm$0.13 & 1.90$\pm$0.31 \\
Cl0340-2840   &       0.068& 0.662&	$2.13^{+0.53}_{-0.20}$     & 3.28$\pm$0.32 & 1.47$\pm$0.14 & 1.90$\pm$0.37 \\
Cl1206-0744   &       0.068& 0.638&	$1.73^{+0.22}_{-0.05}$     & 3.66$\pm$0.27 & 1.78$\pm$0.13 & 1.38$\pm$0.16 \\
A1775         &       0.076& 0.962&	$3.89^{+0.09}_{-0.09}$     &25.64$\pm$0.16 & 9.29$\pm$0.58 &10.60$\pm$1.09 \\
A744          &       0.076& 0.682&	$2.45^{+0.19}_{-0.20}$     & 5.02$\pm$0.29 & 2.21$\pm$0.86 & 1.98$\pm$0.22 \\
RXJ1159+5531  &       0.081& 0.617&	$1.68^{+0.02}_{-0.02}$     & 2.74$\pm$0.07 & 1.43$\pm$0.04 & 1.14$\pm$0.12 \\
Cl2220-5228   &       0.102& 0.890&	$3.79^{+0.48}_{-0.48}$     &15.11$\pm$0.81 & 5.48$\pm$0.59 & 4.64$\pm$0.55 \\
Cl0336-2804   &       0.104& 0.661&	$2.06^{+0.43}_{-0.28}$     & 4.44$\pm$0.48 & 2.02$\pm$0.22 & 2.49$\pm$0.44 \\
Cl1501-0830   &       0.108& 0.745&	$1.93^{+0.14}_{-0.33}$     &13.62$\pm$0.95 & 6.49$\pm$0.45 & 3.88$\pm$0.49 \\
A2220         &       0.111& 0.727&	$2.67^{+0.35}_{-0.27}$     & 7.58$\pm$0.35 & 3.20$\pm$0.15 & 3.83$\pm$0.42 \\
Cl0057-2616   &       0.113& 0.812&	$2.56^{+0.40}_{-0.25}$     &13.42$\pm$0.86 & 5.70$\pm$0.36 & 5.66$\pm$0.68 \\
Cl0838+1948   &       0.123& 0.738&	$3.11^{+0.60}_{-0.59}$     & 5.73$\pm$0.59 & 2.26$\pm$0.23 & 1.96$\pm$0.50 \\
Cl0237-5224   &       0.134& 0.703&	$2.45^{+0.38}_{-0.36}$     & 7.48$\pm$0.69 & 3.27$\pm$0.30 & 3.33$\pm$0.35 \\
Cl1552+2013   &       0.136& 0.603&	$2.52^{+0.67}_{-0.31}$     & 5.12$\pm$0.33 & 2.19$\pm$0.14 & 2.29$\pm$0.28 \\
RXJ1416.4+2315&       0.138& 0.623&	$2.97^{+0.37}_{-0.37}$     &11.28$\pm$0.47 & 5.33$\pm$0.22 & 6.09$\pm$0.64 \\
Cl0245+0936   &       0.147& 0.667&	$2.36^{+0.44}_{-0.44}$     & 4.74$\pm$0.80 & 2.19$\pm$0.37 & 3.41$\pm$1.10 \\
\hline
\end{tabular}
\end{table*}
\setlength{\extrarowheight}{0em}

After this data reduction, the properties of the clusters were
measured following the methods described in \citet{ben12}, and we
recap the key points here. Gas masses were measured by fitting a
projected model of the gas density profile to the observed profile of
the projected emissivity. The gas temperature and luminosity were
measured from spectra extracted within $R_{500}$ (the radius within
which the overdensity is 500 times the critical density at the cluster
redshift). The value of $R_{500}$ was determined iteratively from the
\YM\ scaling relation of \citet{vik09}
\begin{align}
M_{500} = E(z)^{-2/5}A_{YM}\left(\frac{Y_X}{3\times 10^{14}M_\odot keV}\right)^{B_{YM}}
\end{align}
where the $Y_X$ is the product of temperature (measured with the
central $15\%$ of \rf\ excluded) and gas mass, $A_{YM}$ = 5.77
$\times 10^{14}h^{1/2}M_\odot$, $B_{YM}$ = 0.57.

For all analyses, the blank-sky background files were used to estimate
the background level at the cluster position. For the imaging
analysis, which was performed in the $0.7-2\keV$ band, the background
was normalised to match the count rate in parts of the cluster field
that were free from source emission. For the spectral analysis, the
exposure time of the blank-sky files was adjusted so their count rates
matched those of the cluster data in the $9.5-12\keV$ band. A residual
spectrum was produced by subtracting a background-field spectrum from
a source-field spectrum in a region free from source emission. This
residual spectrum was fit with an unabsorbed APEC \citep{smi01} model
with $T=0.18\keV$, which was included as an extra component in all
subsequent spectral fits to the cluster spectra \citep[see][]{vik05a}.

Luminosities and temperatures were measured within \rf\ both with and
without the central $0.15\rf$ of the aperture being included.
Luminosities were measured in the bolometric band and the
$(0.5-2)\keV$ band. All luminosities are unabsorbed, measured in the
cluster rest frame, and are projected luminosities (i.e. not corrected
for the fact that emission beyond \rf\ is projected onto the cluster,
or that the exclusion of the central projected $0.15\rf$ also excludes
emission outside that 3D radius along the line of sight). We use the
notation that \Ttot\ and \Tce\ refer to temperatures measured in the
\rf\ and $(0.15-1)\rf$ apertures respectively, and \Ltot\ and \Lce\
refer to bolometric ($0.01-100\keV$) luminosities in the same
apertures. We use \Lcxo\ to denote the soft band ($0.5-2\keV$)
luminosity in the \rf\ aperture. Finally, we use \Lrosat\ to denote
the soft band ($0.5-2\keV$) luminosity measured with the \ROSAT\ PSPC
in the original 400d survey data (all other luminosities were measured
with \Chandra). The properties of the clusters are summarised in
Tables \ref{ltc} and \ref{lt}.

\setlength{\extrarowheight}{.4em}
\begin{table*}
\caption{\label{lt} Cluster properties measured in the (0.15-1)\rf\
  aperture. \Lce\ is the \Chandra\ bolometric luminosity.}
\centering
\begin{tabular}{lccc}
\hline
Cluster &       $z$ & $\Tce$ (keV) & $\Lce$ ($10^{43}\ergps$) \\
\hline
Cl0327+0233    & 0.030 & $0.92^{+0.07}_{-0.10}$	& $0.18  \pm 0.02$\\
Cl0306-0943    & 0.034 & $1.19^{+0.11}_{-0.12}$	& $0.28  \pm 0.07$\\
Cl1058+0136    & 0.038 & $2.33^{+0.30}_{-0.26}$	& $9.80  \pm 0.88$\\
Cl1259+3120    & 0.052 & $1.00^{+0.07}_{-0.11}$	& $0.42  \pm 0.10$\\
Cl0334-3900    & 0.062 & $2.56^{+0.73}_{-0.52}$	& $1.76  \pm 0.17$\\
Cl0810+4216    & 0.064 & $3.67^{+0.60}_{-0.73}$	& $4.04  \pm 0.38$\\
Cl1630+2434    & 0.065 & $2.55^{+0.52}_{-0.38}$	& $4.06  \pm 0.39$\\
Cl1533+3108    & 0.067 & $1.52^{+0.17}_{-0.23}$	& $2.66  \pm 0.36$\\
Cl0340-2840    & 0.068 & $2.09^{+0.45}_{-0.26}$	& $2.66  \pm 0.29$\\
Cl1206-0744    & 0.068 & $1.72^{+0.14}_{-0.10}$	& $2.88  \pm 0.23$\\
A1775          & 0.072 & $3.65^{+0.13}_{-0.14}$	& $16.60 \pm 0.14$\\
A744           & 0.076 & $2.17^{+0.28}_{-0.16}$	& $3.14  \pm 0.16$\\
RXJ1159+5531   & 0.081 & $1.72^{+0.10}_{-0.04}$	& $1.46  \pm 0.06$\\
Cl2220-5228    & 0.102 & $3.32^{+0.48}_{-0.44}$	& $11.78 \pm 0.77$\\
Cl0336-2804    & 0.104 & $1.86^{+0.29}_{-0.34}$	& $3.68  \pm 0.47$\\
Cl1501-0830    & 0.108 & $1.91^{+0.14}_{-0.33}$	& $13.14 \pm 0.93$\\
A2220          & 0.111 & $2.62^{+0.34}_{-0.30}$	& $6.59  \pm 0.33$\\
Cl0057-2616    & 0.113 & $2.37^{+0.30}_{-0.34}$	& $10.10 \pm 0.80$\\
Cl0838+1948    & 0.123 & $3.26^{+0.82}_{-0.85}$	& $3.79  \pm 0.64$\\
Cl0237-5224    & 0.134 & $2.04^{+0.45}_{-0.32}$	& $5.85  \pm 0.71$\\
Cl1552+2013    & 0.136 & $2.57^{+0.91}_{-0.32}$	& $4.75  \pm 0.51$\\
RXJ1416.4+2315 & 0.138 & $2.72^{+0.35}_{-0.35}$	& $8.13  \pm 0.44$\\
Cl0245+0936    & 0.147 & $2.09^{+0.64}_{-0.35}$	& $3.45  \pm 0.70$\\

\hline
\end{tabular}
\end{table*}
\setlength{\extrarowheight}{0em}

To assess the dynamical state and presence of any cool cores for this
sample, we use the cuspiness of the ICM density profile, the core flux
ratio $F_{core}$, and a visual classification of the X-ray morphology
to categorise clusters. Cuspiness is defined as the logarithmic slope
of the gas density profile at $0.04\rf$, which was measured from our
3D gas density model. $F_{core}$ is defined as the ratio of the
unabsorbed bolometric flux within the $0.15R_{500}$ aperture to the
total flux in the $R_{500}$ region. A cluster was categorised as a
cool core (CC) cluster if it had a cuspiness $>0.7$ and $F_{core}>0.3$
(6/23 clusters). Clusters were classed as being relaxed if the X-ray
images were smooth and symmetric, without secondary peaks or other
substructures (12/23 clusters). With these measurements, we defined a
sample of 5 relaxed cool-core clusters (RCC) as being those relaxed
clusters that were also CC clusters. The other 18 clusters were
classed as non-relaxed-cool-core (NRCC; i.e. the complement of
  the RCC set). The dynamical properties of each cluster are
summarised in Table \ref{nccrcc}.

\begin{table*}
\caption{\label{nccrcc} Dynamical properties and classification of the
clusters.}
\centering
\begin{tabular}{lccccc}
\hline
Cluster          &  Relaxed  & $F_{core}$  & cuspiness  & CC     &RCC     \\
\hline
Cl0327+0233      &           &0.44         &1.02        &$\surd$ &       \\
Cl0306-0943      & $\surd$   &0.59         &1.54        &$\surd$ &$\surd$\\
Cl1058+0136      &           &0.23         &0.70        &        &       \\
Cl1259+3120      & $\surd$   &0.24         &0.49        &        &       \\
Cl0334-3900      &           &0.12         &0.55        &        &       \\
Cl0810+4216      & $\surd$   &0.39         &0.89        &$\surd$ &$\surd$\\
Cl1630+2434      & $\surd$   &0.20         &0.70        &        &       \\
Cl1533+3108      & $\surd$   &0.21         &0.26        &        &       \\
Cl0340-2840      &           &0.19         &0.49        &        &       \\
Cl1206-0744      &           &0.21         &0.94        &        &       \\
A1775            &           &0.35         &0.66        &        &       \\
A744             & $\surd$   &0.37         &0.74        &$\surd$ &$\surd$\\
RXJ1159+5531     & $\surd$   &0.45         &1.31        &$\surd$ &$\surd$\\
Cl2220-5228      & $\surd$   &0.22         &0.54        &        &       \\
Cl0336-2804      &           &0.17         &0.41        &        &       \\
Cl1501-0830      &           &0.04         &0.01        &        &       \\
A2220            &           &0.09         &0.87        &        &       \\
Cl0057-2616      & $\surd$   &0.25         &0.49        &        &       \\
Cl0838+1948      & $\surd$   &0.34         &0.81        &$\surd$ &$\surd$\\
Cl0237-5224      &           &0.22         &0.51        &        &       \\
Cl1552+2013      &           &0.05         &0.76        &        &       \\
RXJ1416.4+2315   & $\surd$   &0.63         &-           &        &       \\
Cl0245+0936      & $\surd$   &0.27         &0.93        &        &       \\
\hline
\end{tabular}
\end{table*}

\subsection{Notes on individual groups}
In some instances the luminosity of a group measured with \Chandra\
significantly differed from that measured with \ROSAT. Some variation
is expected due to differences in calibration and apertures, and is
modelled with a nuisance parameter in our Bayesian analysis. In this
section we discuss groups where the difference in flux is significant,
or where non-standard steps were required in our analysis.

{\em Cl0334-3900:} This group has an irregular morphology and was
classified as a multiple component system by the 400d detection
algorithm. The value of $\Lrosat$ in Table \ref{ltc} is for the main
component. For the \Chandra\ analysis, we included all of the flux
within $\rf$ of the main component, resulting in a higher flux. When
the fluxes from all components in the 400d catalogue are combined, the
\Chandra\ and 400d fluxes agree well. The use of the 400d flux of the
main component alone is consistent with the 400d selection function,
but in fact the detection probability for this group is $\approx1$
whichever flux is used so the choice makes no practical difference in
our analysis.

{\em Cl1501-0830:} This group also has an irregular morphology, with a
clump to the South-West that was excluded in the 400d analysis due to
its proximity to a bright point source. The point source is resolved
and excluded in the \Chandra\ analysis allowing both components of the
group to be included in the \Chandra\ flux. The exclusion of the
South-West clump in the 400d flux has negligible impact on our results
since the 400d detection probability of the group is $\approx1$
regardless, and the difference between the \Chandra\ and 400d fluxes
is absorbed in our nuisance parameter.

{\em A2220:} Part of the emission from this system extended off the
\Chandra\ ACIS-S array. In our standard analysis, any flux missing in
an aperture due to excised sources or chip gas is corrected by assuming
azimuthal symmetry of the X-ray surface brightness profile. In the case
of A2220 this was not appropriate as the emission was significantly
elongated in the direction toward the chip gap (meaning an azimuthal
average would underestimate the missing flux). In this case, the missing
flux was corrected by scaling for the off-chip regions using the flux
distribution in the \ROSAT\ image. This increased the \Chandra\ flux
by $\approx 40\%$.

{\em RXJ1416.4+2315:} The observation of this group exhibited
background flares throughout, and so an extra power-law component was
included when modelling the residual spectrum, as described in the
appendix.

\section{The \LT\ Relation Without Bias Corrections}
\label{sec:bces}
In this section we present the bolometric \LT\ relation for the 400d
groups sample, without accounting for any selection biases.

The correlation between $L$ and $T$ arises in the self-similar model
due to the way that the gas mass, temperature, and cluster structural
parameters all scale with mass, with an additional assumption that the
luminosity is dominated by bremsstrahlung emission \citep[see][for an
expanded discussion]{ben14}. This gives rise to the expectation of a
self-similar bolometric \LT\ relation of the form $L\propto T^2$. For
cooler clusters where line emission becomes comparable to the
bremsstrahlung component, the slope should flatten.

To measure the \LT\ relation, our data were fit with a power-law model
of the form
\begin{align}
\frac{L}{L_0} & = E(z)^{\gLT} \ALT \left(\frac{T}{T_0}\right)
^{\BLT}
\end{align}
For all of our fits, we used $L_0 = 1.0\times 10^{43}\ergps$,
$T_0= 2.0 \keV$, and fixed $\gLT=1$ (for self-similar evolution -- a
negligible assumption for the redshift range covered here).

For the purposes of comparison with other work, and to assess the size
of the sample selection biases, the \LT\ relation was first measured
without modelling the selection biases. The \LT\ data were fit in
base-10 log space using BCES orthogonal linear regression following
\citet{akr96}. The intrinsic scatter of the \LT\ relation ($\intLT$)
was measured by determining the additional error component needed to
give a reduced $\chi^2$ of unity, as in \citet{ben07a}.

The relation was fit for the different apertures used to measure $L$
and $T$, and the best fitting relations are summarised in Table
\ref{lrblt}. In both cases, the slope of the \LT\ relation is
significantly steeper than the self-similar expectation of $\BLT=2$.

\begin{figure}
\begin{center}
\includegraphics[trim=3cm 2.5cm 3cm 3cm, clip=true, width=9cm]{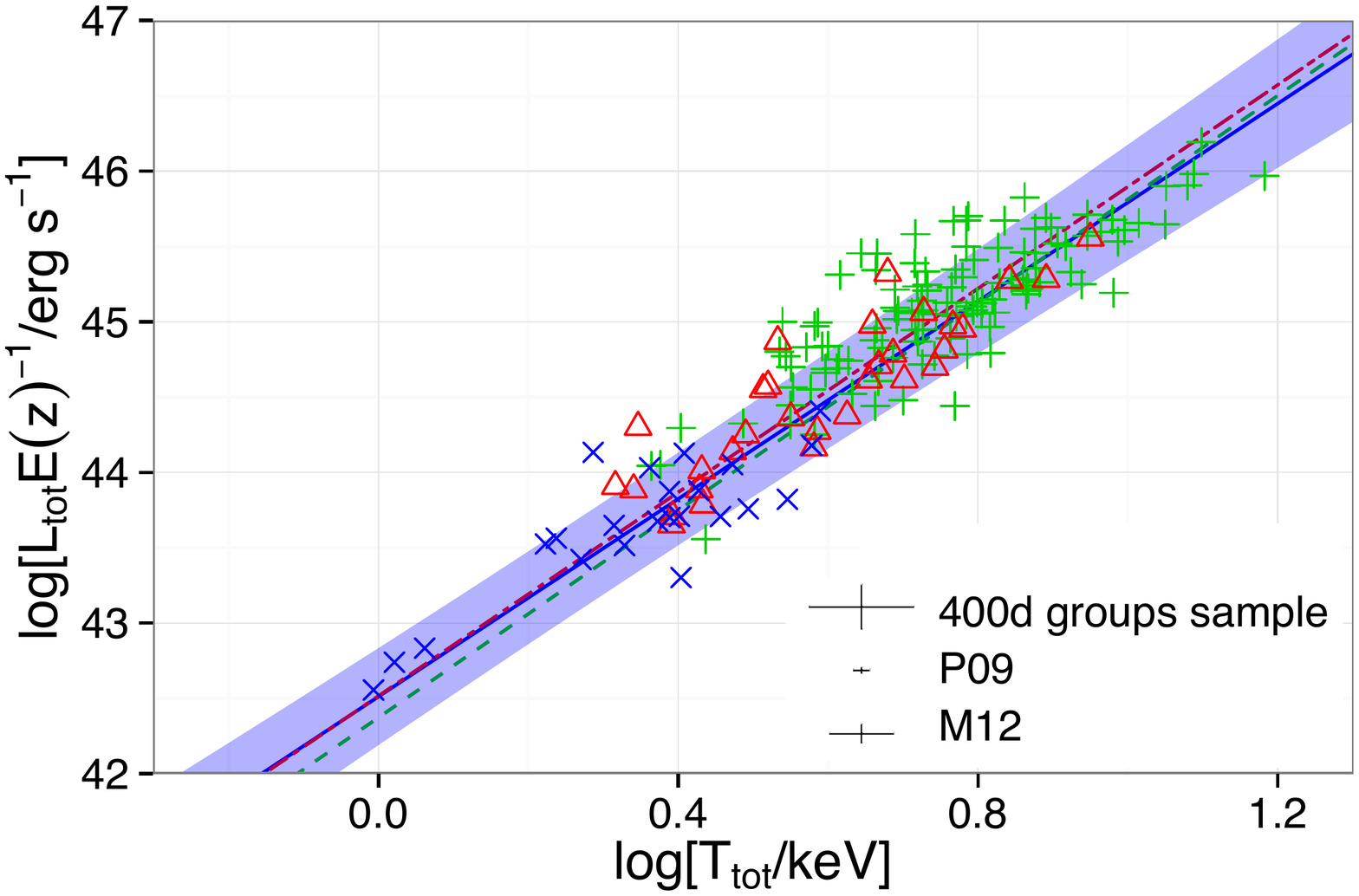}
\\
\includegraphics[trim=3cm 3cm 3cm 3cm, clip=true, width=9.3cm]{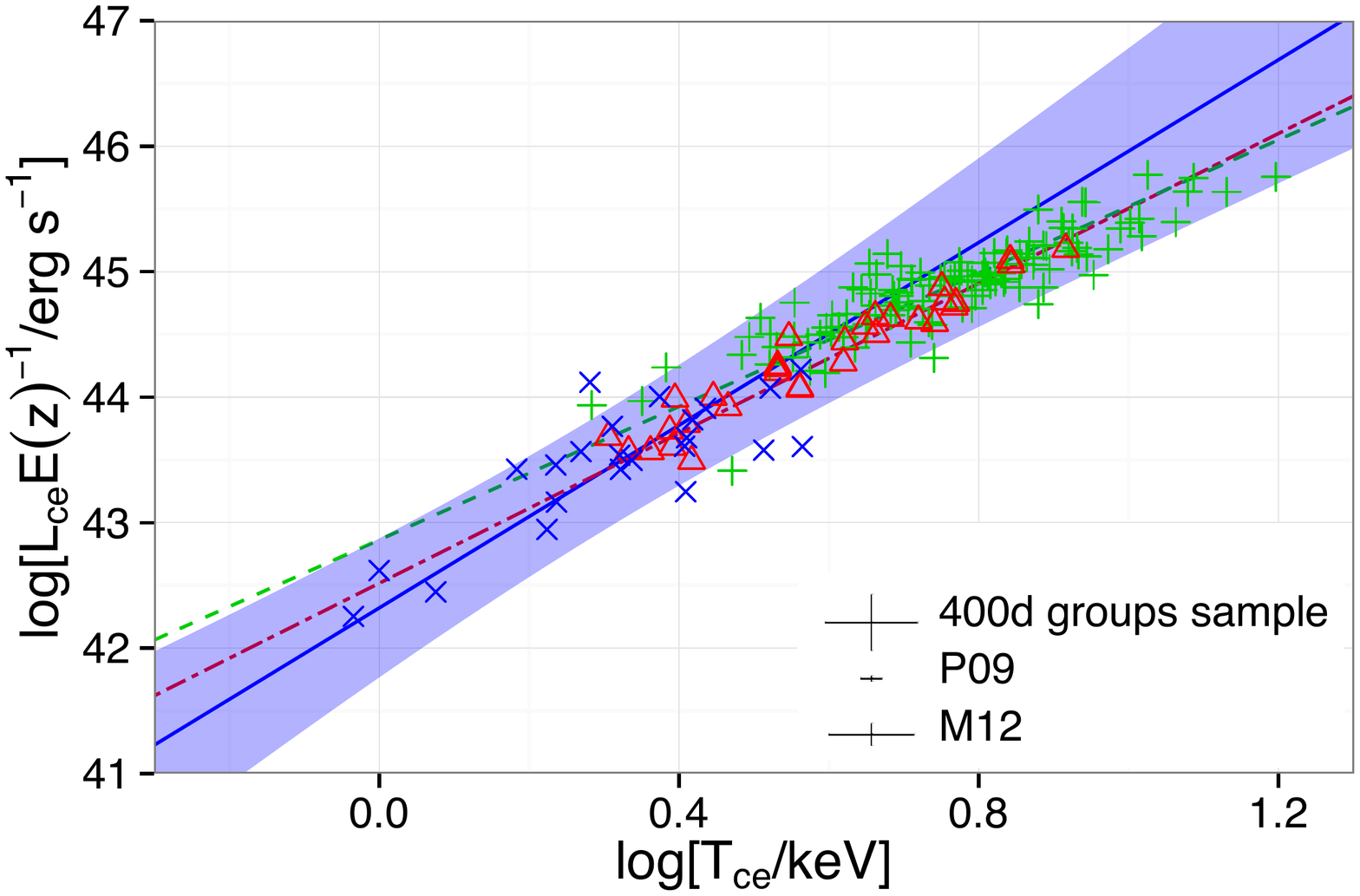}
\end{center}
\caption[]{\label{ltbces} The \LT\ relation (without bias correction)
  for the 400d groups sample (blue crosses, blue solid line with shaded
  error region) is compared with the REXCESS clusters from P09 (red
  triangles and red dot-dashed line), and the archival \Chandra\
  sample of M12 (green plus symbols and green dashed line). {\em Top:}
  quantities measured within the $(0-1)\rf$ aperture. {\em Bottom:}
  quantities measured within the $(0.15-1)\rf$ aperture. In both
  plots, the legend shows representative error bars for the samples.}
\end{figure}

We compared the \LT\ relation for this sample with the previously
published relations of \citet[][hereafter P09]{pra09} which
comprised 31 low-z clusters from the REXCESS sample with \XMM\ data,
and \citet[][hereafter M12]{ben12}, containing 114 clusters over
$0.1<z<1.1$ with \Chandra\ data analysed consistently with the present
work. These comparisons are shown in Figure \ref{ltbces} and for the
moment we simply note the good agreement of the new 400d \LT\ relation
with the samples of more massive systems.

    Some caution should be applied with comparing the \LT\ relation of
    the P09 data with those of the 400d groups and M12 samples, as the
    P09 measurements were based on \XMM\ data while the other two
    samples used \Chandra\ data. There is a significant offset between
    the temperatures measured with the two observatories, with the \XMM\
    temperature being systematically lower than those measured with
    \Chandra\ \citep{sch15}. However the difference
    decreases at lower temperatures and is $<\sim 10\%$  below $4 \keV$ so does
    not strongly impact the comparison between distributions of the
    data points in that regime.

One notable difference between the 400d groups sample and those of P09
and M12 is that unlike the other two samples, the 400d groups and
low-mass clusters show no evidence for a reduction in the intrinsic
scatter of the \LT\ relation when the core regions are removed.
Furthermore, the scatter in the 400d sample with core regions excised
is significantly larger than the equivalent scatter in the P09 and M12
samples. This demonstrates that, while the scatter in the more massive
clusters sampled by P09 and particularly M12 is driven by the core
regions, this is not the case for the lower-mass systems in our 400d
sample. We investigated whether the scatter in the 400d groups sample
had an origin in the dynamical state of the objects by fitting the
\LT\ relation to the subsamples of visually-classified relaxed and
unrelaxed clusters. The sample sizes (12 for relaxed and 11 for
unrelaxed subsamples) were not large enough to allow precise measurements
and while the scatter for the unrelaxed clusters ($0.51\pm0.27$) was
larger than the relaxed clusters ($0.66\pm0.12$), the difference was not significant.

\begin{table*}
  \caption{\label{lrblt} \LT\ relation parameters for the 400d groups
    sample without corrections for selection biases. Best fitting
    parameters are given for the whole sample of 23 clusters, and the
    subsamples of RCC (5 clusters) and NRCC systems (18 clusters).
    Luminosities are bolometric, and the relations are given for
    properties measured with core regions included and excised
    for each subsample. The scatter, $\intLT$ is given as a fractional
    value.}
\centering
\begin{tabular}{c|cc|cc|cc}
\hline
\multirow{2}{*}{Parameters} & \multicolumn{2}{c|}{All} & \multicolumn{2}{c|}{RCC} & \multicolumn{2}{c}{NRCC}\\
\cline{2-7}& $(0.15-1)\,\rf$ & $(0-1)\,\rf$& $(0.15-1)\,\rf$ & $(0-1)\,\rf$&$(0.15-1)\,\rf$ & $(0-)\,\rf$\\
\hline

$A_{LT}$ & 2.52 $\pm$ 0.45 &3.18 $\pm$ 0.42 & 1.39 $\pm$ 0.28 &2.51$\pm$ 0.26 &3.08 $\pm$ 0.56 &3.42 $\pm$ 0.57\\

$B_{LT}$ & 3.81 $\pm$ 0.46 &3.28 $\pm$ 0.33 &2.59 $\pm$ 0.37 &2.15 $\pm$0.17 &3.85$\pm$ 0.52 &3.49 $\pm$ 0.41\\

$\intLT$& $0.59\pm0.17$ & $0.47\pm0.11$ & $0.46\pm0.19$ & $0.18\pm 0.01$ &$0.52\pm0.13$ & $0.49\pm0.15$ \\
  \hline
\end{tabular}
\end{table*}

\section{The \LT\ Relation With Bias Correction}\label{sec:lt-relation-with}
The preceding analysis did not take into account selection effects
induced by the use of an X-ray selected sample of clusters. However
the fact that the 400d groups sample is complete, with a well-defined
selection function means that selection effects can be incorporated
into the analysis. To do this we use a Bayesian approach and computed
the likelihood of the observed properties of the sample for a model
which comprises the cluster temperature function, the \LT\ relation
(parametrised by \ALT\ and \BLT; \gLT was fixed at the self-similar
value), its intrinsic scatter (\intLT, modelled as log-normal) and the
sample selection function (which contains a parameter $\xcal$, defined
below, to model systematic differences between \ROSAT\ and \Chandra).

\subsection{The Likelihood}
We use an improved version of the likelihood model of
\citet[][presented in detail in Pacaud et. al. in prep.]{pac07}. The
likelihood starts with the probability of a cluster in the survey
volume having some temperature $T$. This is given by a mass function
\citep[we used][]{tin08} converted to a temperature function
$\phi(T,z)$ by assuming a fixed mass-temperature (\MT)
relation \citep[we use the bias-corrected relation of][which was
calibrated to \Chandra\ temperatures]{ket14}. Then the probability of
a cluster with temperature $T$ having some luminosity $L$ is given by
the model \LT\ relation including its scatter, $P(L|T,z,\theta)$
(where $\theta$ is our set of model parameters). In order to apply the
selection function, soft-band luminosities are used throughout, and
the selection probability $P(I|L,z)$ is applied to the clusters based
on their \Lrosat. Finally, the likelihood of the observations is
computed using their measurement errors, including the original 400d
luminosity, and the luminosity and temperature measured in the
\Chandra\ observations (here we include a nuisance parameter,
$\xcal\equiv\Lcxo/\Lrosat$ to describe any systematic differences between
the 400d and \Chandra\ luminosities).

The likelihood for cluster $i$ is then
\begin{align}\label{eq.pi}
  P(\Lrosati,\Lcxoi,\hat T_i|z_i,\theta) = \int dT \int d\Lrosat\,
  P(T|z_i) \\
  \times P(\Lrosat|T,z_i,\theta)\, P(\Lrosati|\Lrosat) \nonumber \\
  \times P(\Lcxoi|\Lrosat,\xcal)\, P(\hat T_i|T)\, P(I|\Lrosat,\Lrosati,z_i) \nonumber
\end{align}
where hats indicate observed quantities (we neglect measurement errors
on $z$, so $z\equiv\hat z$).

$P(T|z_i)$ is the prior probability that a cluster at redshift $z_i$
would have a temperature $T$, and is given by the normalised
temperature function:
\begin{align}
P(T|z_i) = \frac{\Phi(T,z_i)}{\int dT \, \Phi(T,z_i)}
\end{align}

The selection function $P(I|\Lrosat,\Lrosati,z_i)$ is composed of two
terms: the full 400d selection function (denoted $P_1$), and the flux
cut used to define the 400d groups sample (denoted $P_2$). $P_1$
depends on the nominal ``true'' 400d flux and not the measured flux,
because the scatter between the ``true'' and measured flux is modelled
in the selection function \citep{bur07}. The flux cut $P_2$ however,
is applied to the measured fluxes, so the selection function can be
written
\begin{align}
P(I|\Lrosat,\Lrosati,z_i) = P_1(I|\Lrosat,z_i)\, P_2(I|\Lrosati,z_i)
\end{align}

The probability in Eq. \ref{eq.pi} must be normalised, and this is
done by dividing it by its integral over the whole observable part of
the $L,T$ plane, which we write as
\begin{align}
C_i = \int d\hat T \int d\Lrosath \int d\Lcxoh\, P(\Lrosati, \Lcxoi,\hat T_i|z_i,\theta)
\end{align}

The final likelihood of the sample is then the
product of this normalised probability over all clusters:
\begin{equation}\label{eq:likelihood}
\lik(\Lrosath,\Lcxoh,\hat T|z,\theta) = \prod_i^{N_{det}} \frac{P_i
  (\Lrosati, \Lcxoi,\hat T_i,|z_i,\theta)}{C_i}
\end{equation}

\subsection{Implementation}
The likelihood function was multiplied by the prior probability
distributions of the parameters. These were uniform for \ALT, \BLT\
and \intLT, while the prior on \xcal\ was log-normal centred on
$\log(\xcal)=0$ with a standard deviation of $0.5$ in natural log
space. The posterior was then sampled using Markov Chain Monte Carlo
(MCMC) with {\em Laplaces
  Demon} \citep{sta16} in $R$
\citep{r14}. Three independent MCMC chains were run, and we checked
that they had converged after any non-stationary parts had been
discarded from the starts of the chains. The parameters posterior
distributions were then computed from the combined chain and are
reported here as the mean and standard deviation of the posterior
samples.

\subsection{Results of bias-corrected \LT\ fit}\label{sec:results-bias-corr}
\begin{figure}
\begin{center}
\includegraphics[clip=true, width=8cm]{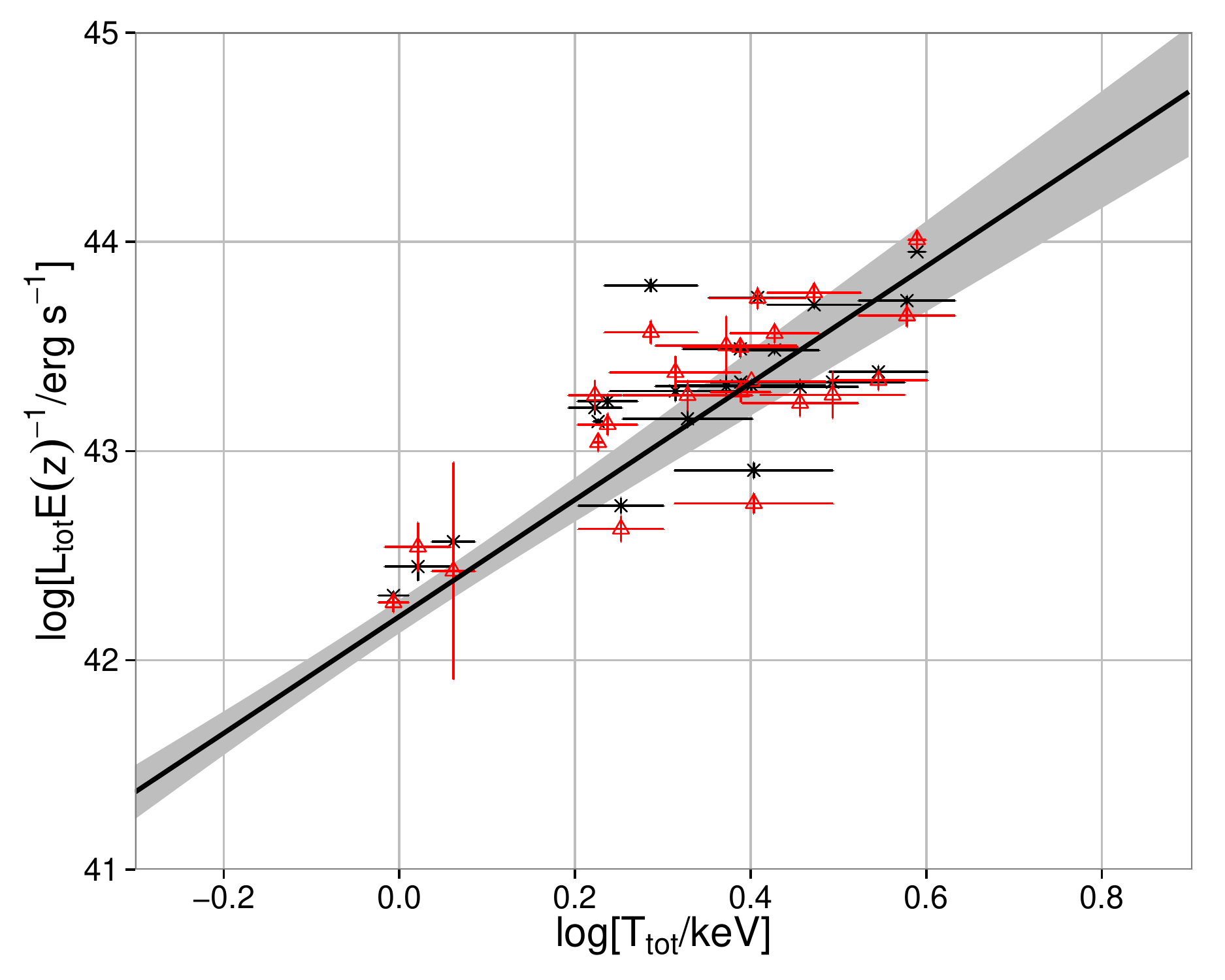}
\end{center}
\caption{\label{fig:crlt}Bias-corrected L-T relation. The black
  crosses are the observed Chandra luminosities, while the red
  triangles show the original 400d ROSAT luminosities (all the
  luminosities are in the $(0.5-2) \keV$ energy band with the core
  regions included). The temperature for each point is the \Chandra\
  temperature from this work.}
\end{figure}

\begin{figure}
\begin{center}
\includegraphics[clip=true, width=8cm]{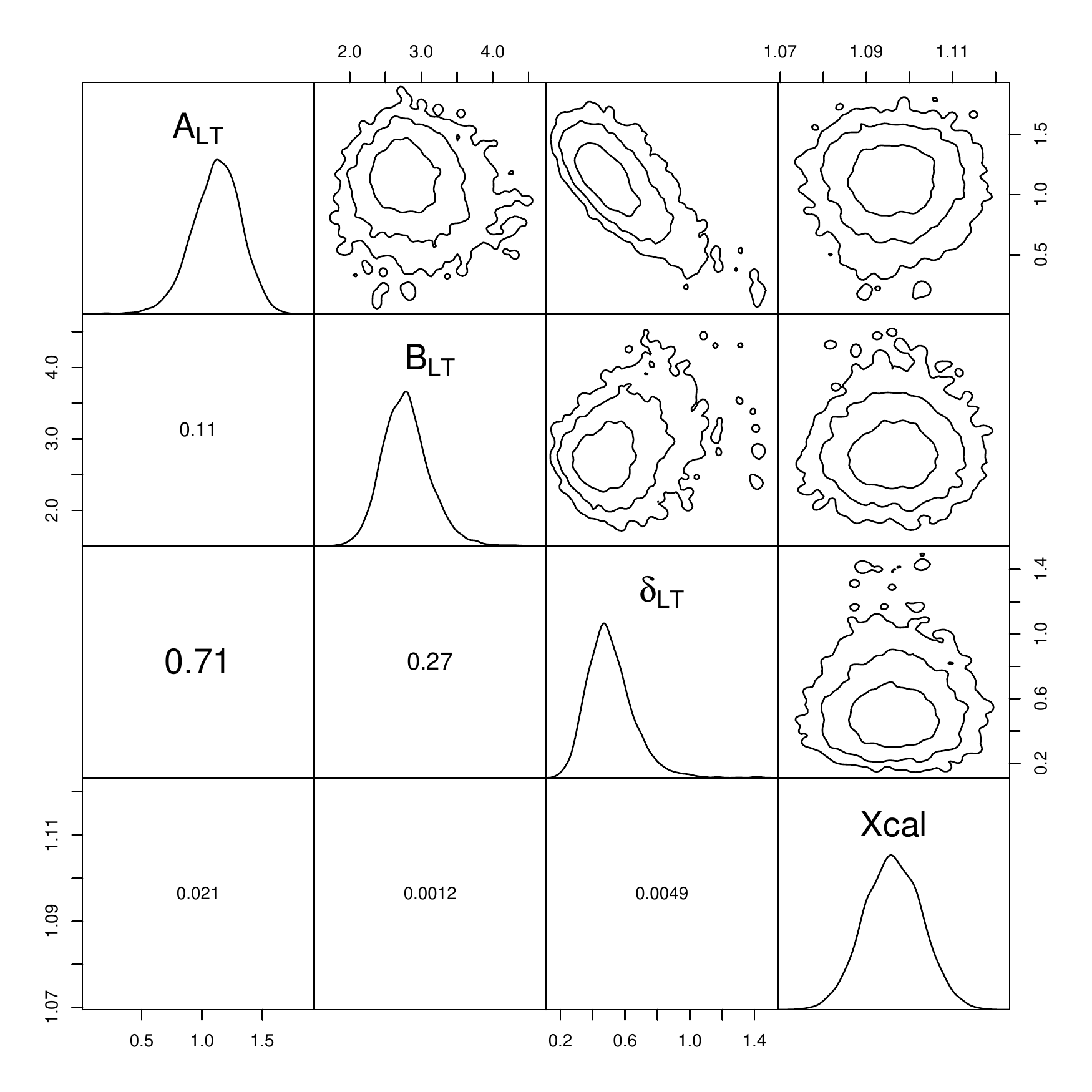}
\end{center}
\caption{\label{fig:crmatrix}Correlation matrix of the bias-corrected
  \LT\ model parameters. The contours shown above the diagonal are
  1$\sigma$, 2$\sigma$ and 3$\sigma$ confidence contours for the
  parameter posterior densities along the diagonal. The values below
  the diagonal are the magnitude of Pearson's correlation coefficient
  for the corresponding pair of parameters, with a font size
  proportional to the strength of correlation.}
\end{figure}

The method described in the preceding sections was used to estimate
the \LT\ relation of the 400d groups sample, including the effects of
selection biases. For these results, we assumed the \MT\ relation of
\citet[][choosing the relation that was bias-corrected and calibrated
to \Chandra\ temperatures]{ket14}; we test the sensitivity of our
results to this assumption later in \textsection
\ref{sec:impact-choice-mt}. The best-fitting bias-corrected relation
is shown in Fig. \ref{fig:crlt}, with \bm{$\ALT=1.12\pm0.21$},
\bm{$\BLT=2.79\pm0.33$}, \bm{$\intLT=0.51\pm0.15$}, and \bm{$\xcal=1.10 \pm 0.01$}.
The parameter posterior distributions are shown in Fig.
\ref{fig:crmatrix}, and are summarised in Table \ref{bcrtable}.

\begin{table}
  \caption[]{\label{bcrtable}Best fitting parameters for the
    $(0.5-2)\keV$ band, core-included \LT\
    relation, including correction for
    selection biases. The columns show the impact of the choice of
    \MT\ relation used in the analysis. The
    first column shows our main results, using the \MT\ relation from
    \citet[][K14]{ket14}. We also show the results for the \MT\
    relations of \citet[][S09]{sun09} and \citet[][F01]{fin01}.
  }
\centering
\begin{tabular}{lccc}
\hline
\MT\          & K14 & S09 &  F01 \\
\hline
$A_{LT}$      & {\bf 1.12$\pm$0.21} & 1.10$\pm$0.21 &  1.11$\pm$0.25 \\
$B_{LT}$      & {\bf 2.79$\pm$0.33} & 2.85$\pm$0.33 &  3.00$\pm$0.35 \\
$\intLT$ & {\bf 0.51$\pm$0.15} & 0.50$\pm$0.15 &  0.53$\pm$0.15 \\
\xcal         & {\bf 1.10$\pm$0.01} &1.10$\pm$0.01 &  1.10$\pm$0.01 \\
\hline
\end{tabular}
\end{table}

\begin{figure}
\includegraphics[clip=true, width=9cm]{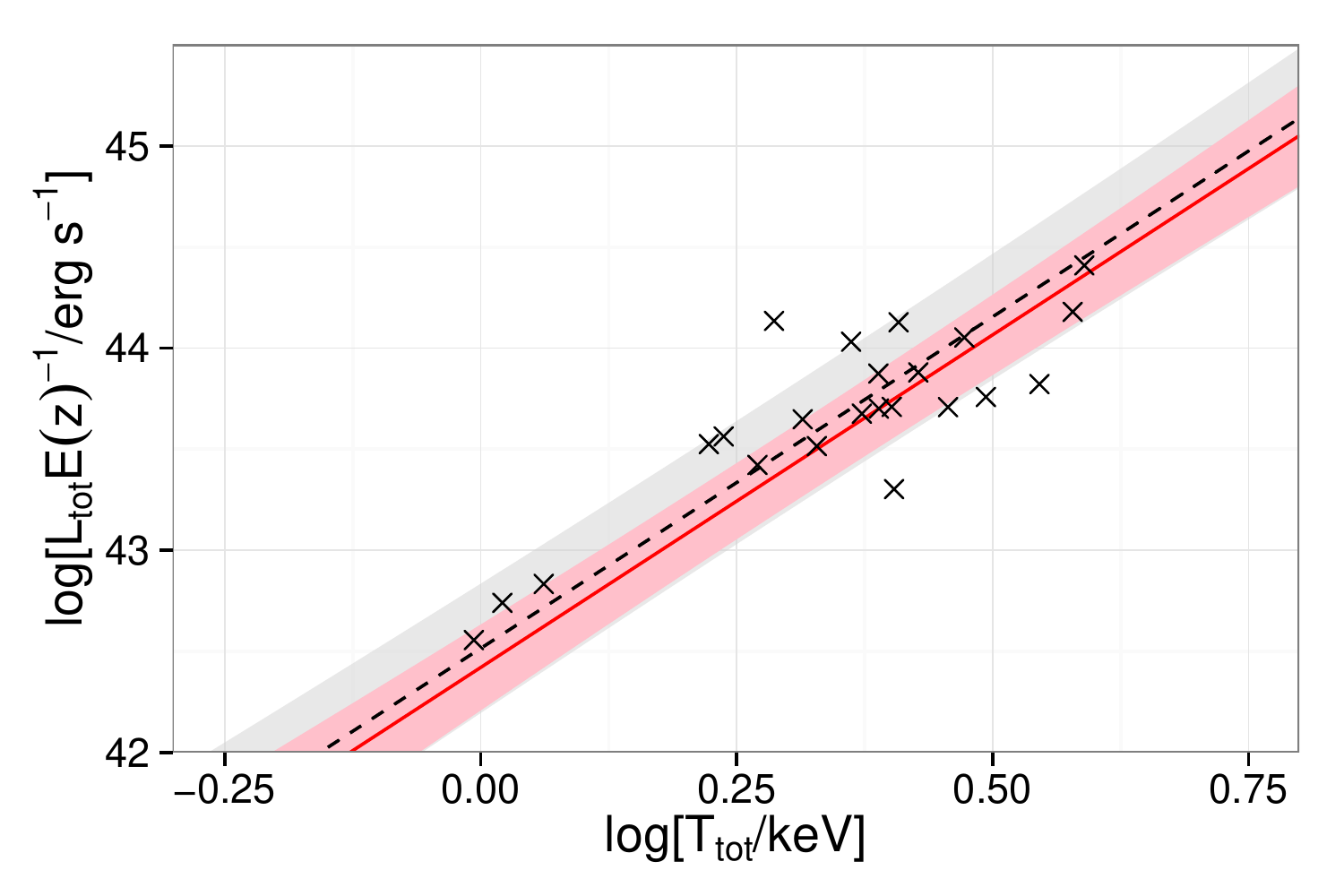}
\caption{\label{fig:biaslt}Bolometric, core-included \LT\ relations of
  the 400d groups sample with and without bias-correction. The black
  dashed line with slightly higher normalisation and grey shaded error
  region is the relation without bias correction. The red solid line
  with pink shaded error region is the bias-corrected model.}
\end{figure}

Recall that for the bias-corrected fit, we use the soft band
luminosities in the $(0-1)\rf$ aperture. In order to compare this fit
with the more widely-used bolometric \LT\ relations, we converted the
bias-corrected soft-band \LT\ relation into a bolometric \LT\ relation
by applying a bolometric correction of the form
\begin{align}
\frac{L_{bol}}{L_{0.5-2}} = 2.3 \left(\frac{T}{2}\right)^{0.50}.
\end{align}
This was determined by generating model APEC spectra in XSPEC
\citep{arn96} and computing the ratios of the fluxes in the
$0-100\keV$
and $0.5-2\keV$
energy bands for models with different temperatures. The ratios as a
function of temperature were fit with a BCES regression in log space
to give the power law parameters. The resulting bias-corrected
estimate of the bolometric \LT\ relation is plotted in Fig.
\ref{fig:biaslt} along with the original bolometric \LT\ relation we
measured in \textsection \ref{sec:bces} without correcting for biases.
The bias-corrected bolometric normalisation was \bm{$\ALT = 2.58\pm0.21$}
and the slope was \bm{$3.29\pm0.33$}.
As expected, the bias correction reduces the normalisation of the
bolometric \LT\ relation, but the size of the effect is small. Neither
the slope or scatter of the \LT\ relation are significantly altered by
our bias correction, indicating that the 400d groups sample is not
strongly effected by Malmquist and Eddington biases.

We also measured the \LT\ relation using the core-excised
temperatures, and the best fitting bias-corrected relation was not
significantly changed. This is consistent with our observation that
the sample does not contain a significant number of strong cool-core
clusters. We proceed with the results based on temperatures measured
with the core regions included due to the improved statistical
precision on the measured temperatures.

As an aside, we note that the absence of strong cool-core clusters in
the 400d groups sample is unlikely to be due to them being rejected by
the 400d determined algorithm. Our simulations have shown that strong
cool core clusters are efficiently recovered by the 400d detection
algorithm even to high redshifts \citep{bur07,vik07}.

\section{Discussion}
\label{sec:discussion}

\subsection{Impact of the assumed \MT\ relation}
\label{sec:impact-choice-mt}
A significant step in our model for the likelihood of the \LT\
observations is the conversion of the theoretical mass function to a
temperature function by way of an assumed \MT\ relation. For our main
analysis we assumed the recent \MT\ relation of \citet{ket14}, which
is calibrated with weak lensing masses, and covers a range of masses
from groups up to clusters. We investigated the sensitivity of our
results to this choice by applying two alternate \MT\ relations from
\citet{sun09} and \citet{fin01}. The best fitting parameters of the
\LT\ relation for the different choices of \MT\ relation are shown in
Table \ref{bcrtable}. It is clear that the choice of \MT\ relation has
no significant impact on our inference of the \LT\ relation
parameters.

  Our analysis also makes the simplifying assumption that there is
  no scatter in the \MT\ relation. For cosmological analyses based on
  cluster number counts, neglecting this scatter can introduce
  significant biases \citep[e.g. ][]{sah09}. For our likelihood model
  for the \LT\ relation, the effect should be smaller. To first order,
  marginalising over scatter in the \MT\ relation would have the
  effect of smoothing the temperature function in the likelihood. This
  would flatten the temperature function somewhat, reducing the level
  of the Eddington bias in a sample. Any neglected mass-dependence of
  the \MT\ scatter would also influence our measurement of the slope
  of the \LT\ relation. Given that the amount of bias in our sample is
  small (e.g. Fig. \ref{fig:biaslt}), we estimate that the impact of
  neglecting the scatter is small, but this remains a systematic
  uncertainty in our analysis.

\subsection{The slope of the \LT\ relation}
\label{sec:crcomparison}
Our analysis of the 400d groups sample is one of several recent studies
to look at the \LT\ relation of low-mass systems that have accounted
for biases. By compiling our results with those other recent
studies, we can build a reliable picture of the nature of the \LT\
relation in the group regime. In particular we focus on the slope of
the \LT\ relation, which is a diagnostic of feedback processes in the
ICM.

L14 analysed a sample of 20 groups with \XMM\ data, which was
statistically complete, allowing the selection bias effects to be
included in the analysis. Their bias correction was determined by
sampling populations of clusters from a mass function, and assigning
luminosities and temperatures from a combination of luminosity-mass
(\LM) and \LT\ scaling relations. Their selection function was then
applied and the scaling relations were fit to the simulated samples to
find the input \LM\ and \LT\ relations that produced the best
agreement with the observed scaling relations. The \LT\ fits were
performed using BCES $Y|X$ regression, giving a slope of
$\BLT=2.05\pm0.32$ which increased to $2.86\pm0.29$ when the bias
correction was applied. We note that the slopes for their \LT\
relation without bias correction range from $2.05-2.76$ depending on
the type of BCES regression used, illustrating the importance of the
choice of fitting method. These fits used luminosities measured in the
$(0.1-2.4)\keV$ band, and temperatures measured with a variable sized
core region removed.

B14 studied a sample of 26 groups with \Chandra\ data, which formed an
incomplete sample, but for which an estimate of the size of the
selection biases was performed. They found a slope of
$\BLT=2.17\pm0.26$ that increased to $3.20\pm0.26$ when the effects of
selection were approximated. These fits used \ROSAT\ luminosities
corrected to the bolometric band and temperatures measured with
\Chandra, and were performed using BCES $Y|X$ regression.

Both L14 and B14 found relatively strong steepening of the \LT\
relation when they included a correction for selection effects, while
for the 400d groups sample, our bias correction made a negligible
change to the slope compared with the original BCES fit. The fact that
we don't see a significant change in the slope when we made the bias
correction appears to be due to the fact that our fit without
correcting for selection biases was performed with an orthogonal BCES
regression, while L14 and B14 primarily used $Y|X$ regression, which
is more sensitive to the selection function since the selection acts
in the $Y$ direction. If we fit our bolometric with a $Y|X$
  regression, the recovered slope is $2.80\pm0.22$, supporting this
  interpretation.

\begin{figure}
\includegraphics[clip=true, width =9cm]{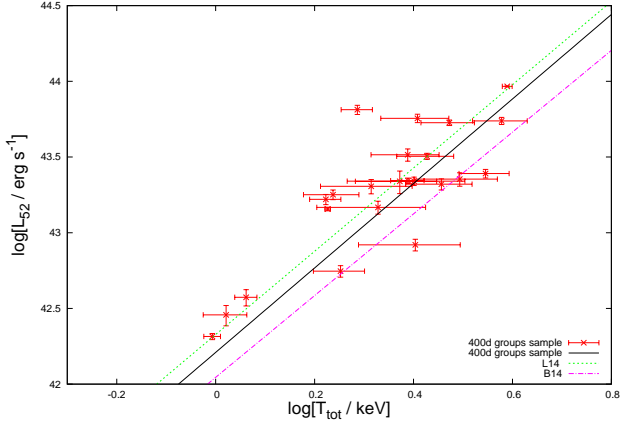}
\caption{\label{fig:biasalllt}The 400d groups bias-corrected L-T
  relation is compared with the bias-corrected fits from L14 and B14.}
\end{figure}

Comparisons between bias-corrected \LT\ relation of the 400d groups
sample and the those of L14 and B14 are presented in Figure
\ref{fig:biasalllt}. For this plot, the L14 and B14 relations were
corrected to the $(0.5-2)\keV$ band using the method described in
\textsection \ref{sec:results-bias-corr}. Both the L14 and B14
relations are consistent with the 400d groups \LT\ relation.

\begin{figure}
\includegraphics[clip=true, width=8.2cm]{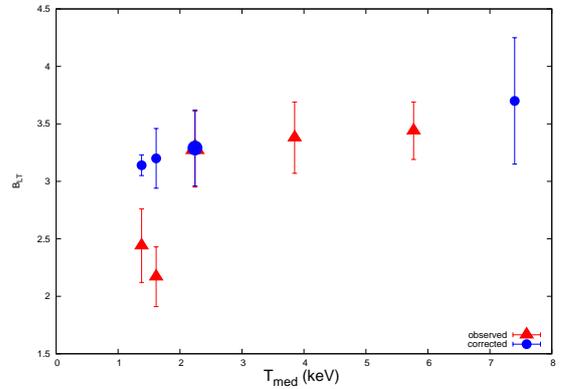}
\caption{\label{fig:bt}The slope of the \LT\ relation (\BLT) is
  plotted against the median temperature ($T_\text{med}$)
  of the sample used to measure the relation for samples summarised in
  Table \ref{bttable}. All of the relations are converted into
  bolometric \LT\ as described in the text. Red triangles represent fits
  where selection biases were not included, while blue circles show
  fits were selection effects were modelled. The points with larger
  symbols at $T_\text{med}=2.2\keV$ are from the present work.}
\end{figure}

We now take advantage of the growing number of studies of the group
\LT\ relation to investigate whether the slope of the \LT\ relation is
steeper for groups than for clusters. In Table \ref{bttable} we
summarise the slopes of the bolometric \LT\ relation measured in
several studies, along with the median temperature of the sample used
to measure the relation. Relations were converted to the bolometric
band as described in \textsection \ref{sec:results-bias-corr}, and
where available we report the bias-corrected slopes. These data are
plotted in Fig. \ref{fig:bt}.

\begin{table}
  \caption{\label{bttable}Summary of \LT\ relation slopes from the
    recent literature. $T_\text{med}$ is
    the median temperature of each sample and BC indicates \LT\
    relations with bias-correction. All the relations are converted
    into bolometric luminosities as described in the main text. The
    rows in bold are from the present work.}
\centering
\begin{tabular}{lcc}
\hline
Sample&    $T_\text{med}$     & \BLT\  \\
\hline
L14                         & 1.4    & 2.44$\pm$0.32\\
L14 (BC)                    & 1.4    & 3.25$\pm$0.09\\
B14                         & 1.6    & 2.17$\pm$0.26\\
B14 (BC)                    & 1.6    & 3.20$\pm$0.26\\
\textbf{400d low-mass}      & {\bf 2.2}    & {\bf 3.28$\pm$0.33}\\
\textbf{400d low-mass (BC)} & {\bf 2.2}    & {\bf 3.29$\pm$0.33}\\
P09                         & 3.9    & 3.38$\pm$0.31\\
M12                         & 5.8    & 3.44$\pm$0.25\\
\citet{man10} (BC)          & 7.4    & 3.70$\pm$0.55\\
\hline
\end{tabular}
\end{table}

Fig. \ref{fig:bt} shows clearly that once selection biases are taken
into account, there is no evidence that the \LT\ relation in groups is
any steeper than for samples of more massive clusters. This
illustrates the importance of correcting selection biases to recover
the true underlying population properties.

Based on the results of \citep{sch15},
in order to compare the slope of an \LT\ relation measured \Chandra\ to one
measured with \XMM\, the slope measured with should be multiplied by $\approx 0.89$
to correct for the calibration differences betwee the instruments.
However, the calibration comparisons made by \citet{sch15}
were almost exclusively limited to temperatures greater than $2 \keV$.
For this reason it is not clear how well their correction would apply
to the low temperature groups in L14. In principal, the slope measured
by P09 could be reduced by an amount roughly the size of the statistical error on
their slope to make it consistent with the \Chandra\ \LT\ relations in
Fig. \ref{fig:bt}, but we chose not to apply any correction to the data.

However, there are additional complexities with interpreting this
result. It is conventional to compare the slope of the \LT\ relation
with the self-similar prediction of $\BLT=2$, but recall that this is
the self-similar slope assuming bolometric bremsstrahlung emission;
for temperatures below about $2\keV$, the contribution of line
emission to the total bolometric luminosity becomes significant. This
effect is well known, but its implication for interpreting the slope
of the \LT\ relation is often overlooked. If the true bolometric
bremsstrahlung \LT\ relation of groups and clusters followed the
self-similar predictions (i.e. there were no feedback effects), then
the observed \LT\ relation would flatten below $\sim2\keV$ due to the
increasing addition of line emission to the self-similar
bremsstrahlung component. Conversely, if the slope of the bolometric
\LT\ relation is observed to be the same for groups and clusters, the
implication is that the increasing contribution of line emission is
masking a steepening of the underlying bremsstrahlung \LT\ relation.

\begin{figure}
\includegraphics[clip=true, width =9cm]{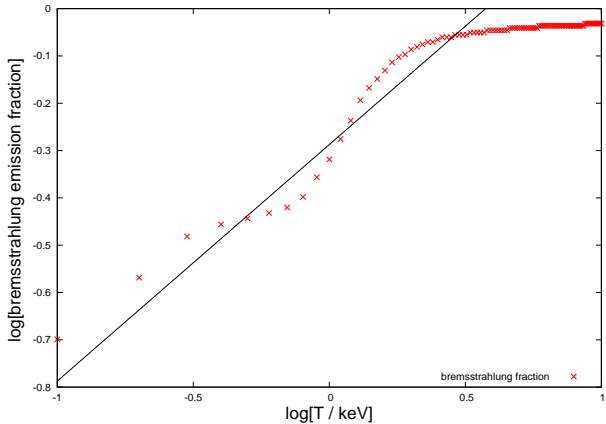}
\caption{\label{bref}Bremsstrahlung emission fraction
  $(L_\text{brem}/L_\text{tot})$ as a function of temperature. This
  illustrates the increasing contribution of line emission to the
  total luminosity for plasmas with $T\lta2\keV$. The solid line is
  our approximate model of the $T<2\keV$ data.}
\end{figure}

We estimated the size of this effect on the \LT\ relation with a
simplistic approach of measuring the luminosity of APEC spectra with a
metal abundance of $Z=0.3$ ($L_\text{tot}$), and then setting $Z=0$
without changing any other parameters to approximate the luminosity of
the pure bremsstrahlung component $L_\text{brem}$. We then used the
ratio $L_\text{brem}/L_\text{tot}$ to approximate the bremsstrahlung
emission fraction for a range of temperatures, and the results are
presented in Figure \ref{bref}. The increasing contribution of line
emission to the total luminosity is clear.

The declining contribution of bremsstrahlung to the total luminosity
for $T<2\keV$ was crudely approximated with a power law of the form
\begin{align}
\frac{L_\text{brem}}{L_\text{tot}} = 0.73 \left(\frac{T}{2}\right)^{0.50}
\end{align}
which is plotted in Fig. \ref{bref}. We thus estimate that for samples
with significant numbers of systems below $2\keV$, the observed
  bolometric \LT\ relation could be steeper than the underlying
  bremsstrahlung \LT\ relation by up to 0.5.

Taking this effect into account, as a first approximation, the
\BLT\ values for the 400d, B14 and M14 samples in Fig. \ref{bref}
should all be raised by $\sim0.5$ in order to assess the impact of any
feedback. With this extra steepening, the slopes of the group \LT\
relations would remain consistent with the cluster \LT\ relations
given the precision of the current measurements, but the effect is
large enough that it should be considered in studies of the group \LT\
relation.

It would be useful to directly measure the bolometric bremsstrahlung
\LT\ relation from the cluster data to make more direct comparisons
with the self-similar model. However, this introduces significant
complications in the modelling of selection biases. This is because
the clusters are detected on the basis of their total emission
(including line emission) so the selection function must be expressed
in those terms. Furthermore, clusters are detected in soft band X-ray
imaging, where the contribution of emission lines is even stronger
than in the bolometric band. This means that for groups, variations in
the metal abundance between systems (perhaps related to their feedback
history) could significantly impact the selection function. This makes
the problem difficult to unpick observationally, and greater success
should result from comparisons of complete X-ray selected group
samples with the output of cosmological hydrodynamical simulations,
onto which the various observational effects can be applied.

The overall picture that emerges from the comparison of these samples
is that the group \LT\ relation is consistent with that of clusters,
and that the bolometric bremsstrahlung \LT\ relation has a slope of
$\approx3-3.5$ across the full range of group and cluster samples;
always significantly steeper than self-similarity.

\subsection{Scatter in the \LT\ relation}
In addition to the slope of the \LT\ relation, its intrinsic scatter
is a signature of the astrophysical processes affecting the ICM in
groups and clusters. Here we will compare the scatter measured for the
400d groups sample with measurements from other group and cluster
samples. In all cases, the scatter is modelled as lognormal and is
reported as the intrinsic scatter in $L$ in natural log space, so
corresponds to a fractional scatter. Unless otherwise stated, the
scatter values are for core-included luminosities.

In our analysis of the 400d groups sample, we found a scatter in the
bias-corrected soft-band \LT\ relation of $\intLT=0.51\pm0.15$. For
the bolometric \LT\ relation without correction for biases, we found
{\bf $\intLT=0.59\pm0.17$}, so the scatter was not significantly affected by
selection biases. The only other measurement of scatter in the group
regime to include an estimate of selection biases was that of B14.
They found a scatter of $\intLT=0.55$ which increased to $\intLT=0.73$
when they approximated the removal of selection biases (errors were
not reported, so it is not clear if the change is significant). B14
reported that the bias-corrected scatter in their group sample was
larger than that in the cluster sample they used ($\intLT=0.62$),
however (as B14 noted), this conclusion is limited by the fact that
their group sample is incomplete and suffered from archival biases.
This could be a significant problem, since the ``interesting'' groups
in the archives are likely to be among those showing the greatest
deviations from the average properties.

The present analysis of the 400d groups sample is the most robust
attempt thus far to measure the scatter in the \LT\ relation including
bias-corrections. The scatter of $0.51\pm0.15$ we find in the group
population is not significantly different from the scatter found for
the cluster population in the bias-corrected measurements of B14
($0.62$) and \citet{man10} ($0.61\pm0.15$).

\subsection{Steepening in the RCC \LT\ relation}
\label{sec:ncrcomparison}

In M12, we found that the bolometric \LT\ relation for the subset of
the 21 most relaxed, cool-core (RCC) clusters in our archival
\Chandra\ sample had a self-similar slope ($\BLT=1.90\pm0.14$) with
negligible scatter, when the central $0.15\rf$ was excluded. We found
a suggestion that the slope of this core-excluded RCC relation might
steepen below about $3.5\keV$ but lacked sufficient numbers of
low-mass systems to make a clear measurement. With the addition of the
400d groups sample, we can extend this test to these lower-mass
objects.

\begin{figure}
\includegraphics[clip=true, width=9cm]{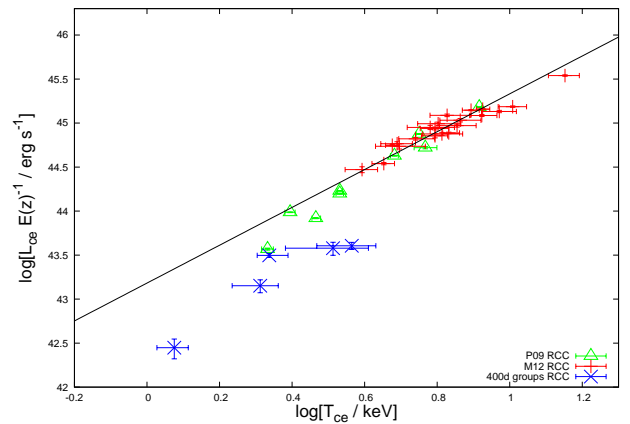}
\caption{\label{fig:ccnocore} The bolometric \LT\ relation for RCC
  clusters from M12 and the 400d groups sample, and CC clusters from
  P09. Luminosities and temperatures were measured in the
  $(0.15-1)\rf$ aperture. The solid line is the best fit to the M12
  data.}
\end{figure}

In Fig. \ref{fig:ccnocore} we plot the 5 RCC clusters from our sample
along with the RCC clusters from M12 and the CC clusters from P09. The
400d points are clearly inconsistent with an extrapolation of the
self-similar M12 relation, but along with the lower-temperature P09
clusters, the data strongly suggest a steepening of the \LT\ relation
below about $3\keV$. Recall that the 400d groups sample has been
analysed in a manner consistent with M12, and the core-included \LT\
relations agree well (Fig. \ref{ltbces}). Some evidence for steepening
in the \LT\ relation is seen in the {\em core-excised} plot in Fig.
\ref{ltbces}, but this is more apparent in the RCC subsample shown in
Fig. \ref{fig:ccnocore}, primarily because of the much flatter
self-similar slope of the M12 RCC clusters.

Putting these results together with the previous section, we find that
when the core regions of clusters are included, the \LT\ relation of
groups is consistent in every way with that of clusters. However, when
core regions are removed, and particularly for the most relaxed,
cool-core clusters, the slope of the \LT\ relation steepens from a
self similar value in the cluster population (M12) to a steeper value
below $\sim3\keV$.

\section{Summary}
\label{sec:summary}
This work presented the \LT\ relation of 23
low-mass galaxy clusters and groups selected from the 400d survey. The
systems were all observed with \Chandra, and the analysis of the \LT\
relation included a Bayesian modelling of the selection biases. The
main results of this work were:

\begin{itemize}

\item The effect of selection biases on the 400d groups sample is not
  large; our correction for the selection biases did not significantly
  change any of the \LT\ relation parameters.

\item The core-included \LT\ relation of the 400d groups sample was
  consistent with the \LT\ relations found in clusters, and
  significantly steeper than self-similar predictions. Indeed, when
  combined with other recent studies of the \LT\ relation in groups,
  and once selection biases are corrected, there is no evidence that
  the slope of the (core included) group \LT\ relation is different
  from that of massive clusters.

\item The intrinsic scatter of the \LT\ relation (with cores included)
  of the 400d groups is $\approx 50\%$ and is consistent with the
  scatter in the \LT\ relation found in cluster samples.

\item While the magnitude of the scatter in the 400d groups \LT\
  relation is consistent with that found for more massive clusters, it
  is not driven by the luminosity of the core regions in the same way
  as for cluster samples. Instead, the scatter in the 400d groups
  relation seems to be driven by the dynamical state of the clusters.

\item The increasing contribution of line emission to the luminosity
  of lower-temperature systems means that the bolometric \LT\ slope
  measured in the group regime is flatter by $\approx0.5$ than the
  underlying bolometric bremsstrahlung \LT\ relation slope. The latter
  is what is predicted by the self-similar model, so this effect could
  (partially) mask processes that are removing gas from lower mass
  systems. In the current study, taking this effect into account would
  not change the conclusion that the slope of the \LT\ relation is
  consistent for groups and clusters.

\item For the particular case of relaxed cool-core (RCC) systems, we
  find that the 400d groups lie significantly below the self similar
  {\em core-excised} \LT\ relation found for massive RCC clusters.
  This suggests a significant steepening of the core-excised RCC \LT\
  relation below about $3\keV$.

\end{itemize}

Overall, our work is the most rigorous attempt so far to measure the
\LT\ relation in the group regime, including the correction for
selection biases. These results thus provide a secure basis against
which to test feedback models in hydrodynamical simulations.
  Those models are often tested against data for which selection
  biases have not been modelled \citep[e.g.][]{sho10,leb14}. Studies
  like ours provide corrected scaling relations that can be compared
  directly with the simulations. For future work, an improved analysis
  would relax the assumption of a fixed \MT\ relation with no scatter
  and instead use a multivariate analysis to model the luminosity,
  temperature and M$_{gas}$ of low-mass clusters simultaneously
  \citep[as in e.g.][]{man10,ben14,evr14,ett15}. However, this would
  require an observable that was a direct proxy for cluster mass, such
  as weak lensing mass, which is currently very challenging to obtain
  at these masses.

\section*{Acknowledgements}
BJM and PAG acknowledge support from STFC grants ST/J001414/1 and
ST/M000907/1.\\RB was supported by RNF grant 14-22-00271.

\bibliographystyle{mnras}

\bibliography{./Master}
\appendix

\section{Analysis of RXJ1416.4+2315}
RXJ1416.4+2315 (obsid = 2024) is an unusual case, with an
unrealistically high-temperature (up to $\sim20\keV$) in our initial
analysis. Upon inspection of the lightcurve and background spectrum,
the observation was found to be affected by low-level background
flaring throughout the observation. For this reason, a bespoke
analysis was performed.

\begin{figure}
\begin{center}
\setlength{\unitlength}{1in}
\scalebox{0.325}{\includegraphics[trim=0cm 1.5cm 1cm 2cm, clip=true]{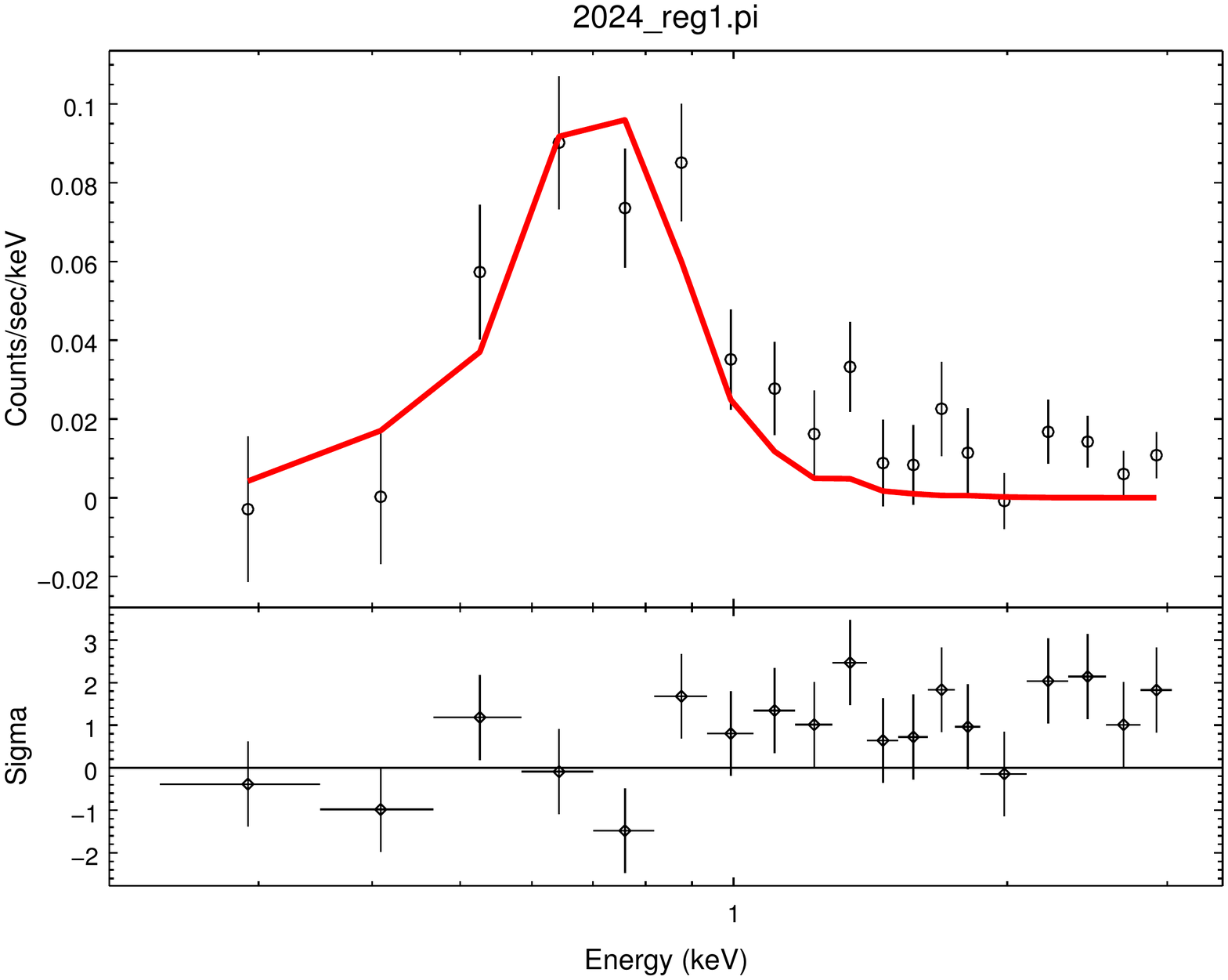}}
\end{center}
\caption{\label{2024a}\small{The background residual spectrum for
    cluster RXJ1416.4+2315 is shown with the best-fitting APEC thermal
    model. Residuals in units of $\sigma$ are shown in the bottom
    panel.}}
\end{figure}

\begin{figure}
\begin{center}
\setlength{\unitlength}{1in}
\scalebox{0.325}{\includegraphics[trim=0cm 1.5cm 1cm 2.5cm, clip=true]{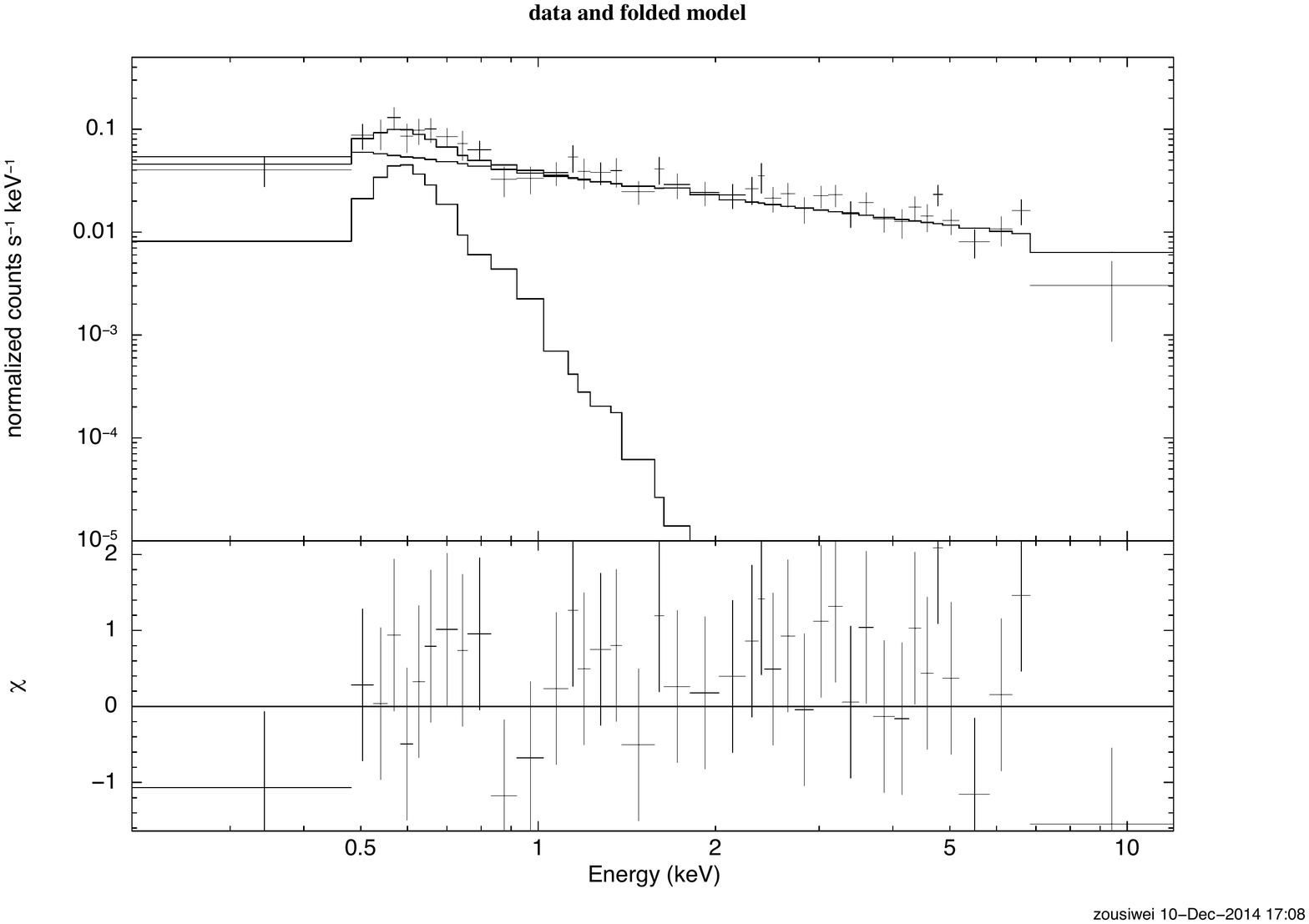}}
\end{center}
\caption{\label{2024b}\small{The background residual spectrum for
    cluster RXJ1416.4+2315 is shown with the best-fitting APEC plus
    power-law model. Residuals in $\chisq$ are shown in the bottom
    panel.}}
\end{figure}

We produced a residual background spectrum of RXJ1416.4+2315, as for
our analyses of the other targets, but in this case it was modelled
with the standard thermal component plus an additional power-law
component. Here, the power-law models the extra particle-induced
background in this observation, and so was not folded through the
instrument effective area. Figure \ref{2024a} shows the residual
spectrum modelled by a thermal model alone (our standard analysis);
the residuals are systematically high above $\sim 1\keV$. Figure
\ref{2024b} shows the same residual background spectrum modelled with
an additional power-law component. This background model was then
included as an extra additive model when fitting the cluster spectrum.
We included an extra systematic component to the errors on the
temperature and normalisation of the cluster APEC model by fixing the
slope and normalisation of the extra power-law component at their
$\pm1\sigma$ errors and refitting the cluster thermal component. The
maximum change in the cluster temperature and normalisation were used
as an estimate of their systematic uncertainty due to the extra
background component. These systematic errors were very similar in
size to the original statistical errors, and were added in quadrature
to the statistical uncertainties on the derived cluster properties.

\end{document}